# An Empirical Study on Collaborative Architecture Decision Making in Software Teams

Sandun Dasanayake, Jouni Markkula, Sanja Aaramaa, Markku Oivo

M3S, Faculty of Information Technology and Electrical Engineering, University of Oulu, Oulu, Finland
{sandun.dasanayake, jouni.markkula, sanja.aaramaa, markku.oivo}@oulu.fi

**Abstract.** Architecture decision making is considered one of the most challenging cognitive tasks in software development. The objective of this study is to explore the state of the practice of architecture decision making in software teams, including the role of the architect and the associated challenges. An exploratory case study was conducted in a large software company in Europe and fifteen software architects were interviewed as the primary method of data collection. The results reveal that the majority of software teams make architecture decisions collaboratively. Especially, the consultative decision-making style is preferred as it helps to make decisions efficiently while taking the opinions of the team members into consideration. It is observed that most of the software architects maintain a close relationship with the software teams. Several organisational, process and human related challenges and their impact on architecture decision-making are also identified.

## 1 Introduction

Software architecture serves as the intellectual centrepiece that not only governs software development and evolution but also determines the overall characteristics of the resulting software system [1]. It provides support for various aspects of software system development by facilitating functions such as enabling the main quality attributes of the system, managing changes, enhancing communication among the system stakeholders and improving cost and schedule estimates [2]. Architecture decisions stand out from the rest because they dictate all downstream design choices; thus, they have far-reaching consequences and are hard to change [3]. Making the right architecture decisions, understanding their rationale and interpreting them correctly during software system development are essential to building a system that satisfies stakeholder expectations. As the system evolves, making new architecture decisions and removing obsolete ones to satisfy changing requirements while maintaining harmony with the existing decisions are crucial to keeping the system on course [4].

A new perspective on software architecture and making architecture decisions has emerged with the popularity of lean and agile development practices [5]. The discussion regarding the big upfront design and continuous design, challenges to find the right balance of initial architecture design and its evolution during the software system life cycle [6]. At the same time, the emphasis on collaboration and agility causes architects to rethink making decisions from their ivory towers [7]. In most cases, architects are now part of the software team, and the important architecture decisions are made by the team rather than an individual architect [8]. With this change of perspective, software architecture decision making is now mostly considered a group decision-making (GDM) activity [9, 10].

## 2   Background

Although the importance of architecture decisions has long been recognised, they only began to gain prominence in software architecture about a decade ago [4]. Since then, architecture decisions and the rationale behind them have been considered first-class entities. Reasons such as dependencies between decisions, considerable business impact, possible negative consequences and a large amount of effort required for analysing alternatives are also recognised as factors contributing to the difficulty of architectural decisions [8]. Due to the importance and complexity of architecture decision making, the research community has given considerable attention to the topic, and a number of techniques, tools and processes have been proposed to assist in different phases of the architecture decision-making process [2]. Even though some attempts have been made to develop GDM solutions for architecture decision making, most of the solutions, including the most widely used ones, are not developed from a GDM perspective [11].

The groups can choose different decision-making methods such as consensus decision making, majority rule, decisions by an internal expert and decisions by an external expert, to reach a decision [12]. Based on the interaction between the team leader and the team, the decision styles in teams can also be classified into many different categories [13–15]. GDM has advantages such as increased acceptance, a large amount of collective knowledge, multiple approaches provided by the different perspectives and better comprehension of the problem and the solution [16]. At the same time, there are also some weaknesses that undermine the use of GDM in certain situations. Liabilities such as being time-consuming and resource heavy, vulnerability to social pressure, possible individual domination and the pursuit of conflicting secondary goals can result in low-quality compromised solutions [16]. One of the major weaknesses of GDM is *groupthink* [17], where the group makes faulty decisions without exploring the solutions objectively because of the social pressure to reach a consensus and maintain the group solidarity.

## 3   Case Study

In this research, the case study approach was selected for two main reasons. First, the case study is recommended for the investigation of a phenomenon when the current perspectives seem inadequate because they have little empirical evidence [18]. Although generic GDM is a well-researched area, few empirical studies have been made about GDM in software architecture. Second, in the case of decision making, the context in which the decision is made is essential to understanding the decision fully [19]. Since the case study allows us to study a phenomenon in its natural setting, the case study makes it possible to gather insights about the phenomenon itself as well as its interactions with its surroundings.

### 3.1 Case Study Design

This exploratory case study was designed to seek new insights into architecture decision making in software teams. A European software company was selected as the case company and the software teams in the company were used as the unit of analysis of the study. The case company specialises in providing software products and services to the consumer market, enterprise customers and third-party service providers. It has around a thousand employees and has a strong global customer base as well as offices and partners around the world. The company's product development activities are carried out in development centres located in multiple countries. At the time of the study, the company had three parallel business units: independent profit centres (BU1, BU2 and BU3) that focused on different product and service offerings, and market segments. In addition, there was a horizontal unit (BU4) that provided common solutions such as backend services for the other thee business lines. Finally, there was a centralised technical decision-making body, the tech committee (TC), which made company-wide technical policy decisions.

Two research questions (RQs) were derived based on the objectives of the study. While the questions are correlated to each other, each question is designed to find answers to a different aspect of the problem.

**RQ1.** *How do software teams make architecture decisions?* The aim of this question is to understand the state of the practice of architecture decision making in the case company, including the processes, tools and techniques, together with the contextual information. Answers to this question will help in understanding the overall architecture decision-making approach of the company as well as the architecture decision-making approaches of individual software teams.

**RQ2.** *What are the challenges in architecture decision making in software teams?* Identifying various challenges faced by the software teams during architecture decision making is the main goal of this research question. Answers to this question will also reveal the underlying sources of those challenges and their impact on architecture decision making.

### 3.2 Data Collection

Fifteen software architects from the different teams of the case company were selected to represents their respective teams (ST1–ST15). As shown in Table 1, they represent all the business and technical units of the company. Despite the variation in the job titles, all of them perform duties as software architects in their respective teams. The software architects are located in three different sites: the headquarters (HQ) and two development centres (DC1 and DC2).

**Table 1.** Interviewee information

| Unit | Team ID | Site | Interviewee Title | Team Size |
|------|---------|------|-------------------|-----------|
| BU1  | ST1     | DC1  | Domain Architect  | 5 |
|      | ST2     | HQ   | Lead Software Engineer | 7 |
|      | ST3     | HQ   | Lead Software Engineer | 4 |
| BU2  | ST4     | DC1  | Domain Architect  | 4–6 |
|      | ST5     | DC1  | Software Architect | 5 |
|      | ST6     | HQ   | Lead Architect | 8 |
|      | ST7     | HQ   | Program Lead | 5–7 |
|      | ST8     | HQ   | Senior Software Engineer | 6 |
| BU3  | ST9     | HQ   | Senior Software Engineer | 8 |
|      | ST10    | DC2  | Domain Architect | 4–7 |
|      | ST11    | HQ   | Software Engineer | 4 |
|      | ST12    | DC2  | Senior Software Engineer | 5 |
| BU4  | ST13    | HQ   | Senior Software Engineer | 7 |
|      | ST14    | HQ   | Senior Software Engineer | 8 |
| TC   | ST15    | HQ   | Chief Architect | N/A |

A set of questions divided into different themes was used to guide the interviews. The interview begins with questions related to the context and then gradually focuses on software architecture and architecture decision making. The interview questions later discuss the challenges that are faced and the possible solutions to these challenges. The interviews were conducted by two researchers. Most of the interviews were carried out face to face on site. Skype was used for three interviews due to travelling and scheduling issues. Each interview lasted about 1.5 hours. All interviews were recorded with the consent of the interviewees.

### 3.3 Data Analysis

A set of decision-making styles derived from the research literature [13–15] was adapted to the software architecture decision-making context to analyse the decision making in the software teams in the case company. Each of these decision-making styles has different characteristics in terms of the decision maker, the origin of the solution and participation in the decision-making process, as shown in Table 2.

**Table 2.** Software architecture decision-making styles

| Decision Style | Decision Maker | Solution Origin | Participation | |
|---|---|---|---|---|
| | | | SW Team | Architect |
| Authoritative | Architect | Architect | Passive | Active |
| Persuasive    | Architect | Architect | Active  | Active |
| Consultative  | Architect | Shared    | Active  | Active |
| Consensus     | Shared    | Shared    | Active  | Active |
| Delegative    | SW Team   | SW Team   | Active  | Passive |

Based on the degree of involvement of each party, these decision styles can be placed on a continuum and grouped into three categories: architect driven (authoritative, persuasive), team driven (delegative) and collaborative (consultative, consensus). In addition to using the above classification to capture the decision-making styles, the identified challenges are categorized into three different groups as organizational, process and human, based on the origin of the challenge.

## 4   Architecture Decision Making in the Case Company

It is clear that most of the software teams in the company follow GDM to make architecture decisions. The decision-making process appears to be informal. However, each team have some form of structured decision-making practice as all the interviewees were able to describe it during the interviews. The software architecture decision-making process in the case company is mainly a two-fold process composed of team level and organisational level decision making. In addition to that, there is also individual level decision-making, as each decision-maker makes individual decisions while participating team level or organizational level decision-making sessions. Even though software teams have freedom to make architecture decisions regarding their own software components, architecture steering groups and the TC are involved in making high-impact decisions that can affect the other teams or the company's business performances.

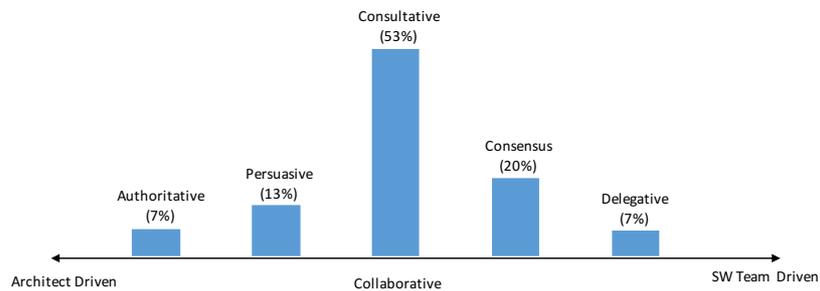

**Fig. 1.** Primary decision making style in software teams

Architecture decision-making styles in each software team are based on the preferences of the software architect and the team members. However, all the interviewees made it clear that they selected the decision-making style based on the context, since there is no "one size fits all" kind of solution. Meanwhile, the decisions related to tasks that have an impact beyond the scope of the team are escalated to the architecture steering groups or the TC. Fig. 1 shows the most commonly used architecture decision-making style of each team. According to that, consultative decision-making is the most commonly used decision-making style; 8 teams (53%) claimed to use that as their primary decision-making style. One notable fact that is brought up during the interviews was the majority of the consultative decision-making style followers are willing to reach consensus during the consultation process if possible. However, they keep consultative decision-making as the primary decision-making approach as it allows them to avoid deadlocks and make timely decisions as the projects demand.

The interviewees provided arguments for choosing and not choosing each decision-making style. The arguments in favour of collaborative decision-making styles are that they increase team motivation, promote continuous knowledge sharing and identify team members who have expertise in the problem domain. The main arguments against these styles are that they are time-consuming and that it is difficult for team members to come to an agreement. Clarity of responsibility and saving time and money were given as reasons for using architect-driven decision-making styles. Others claimed that architecture decision making is too complex to be handled by one person. It can limit the creativity of the solutions and introduce bias into the decisions since all the interviewees use personal characteristics such as experience and intuition for individual level decision-making. The only reason given for opting for delegative decision-making style is that the architect's unwillingness to take the responsibility of the design process.

The consultative decision-making style, which is preferred by the majority, brings the right balance into the decision-making process as it allows the software teams to makes decisions promptly while taking the opinion of the team members into consideration. This style makes it easier to attribute a certain decision to the decision-maker, hence maintaining the design rationale to some extent. The consultation process also helps to share information and spread the knowledge within the team. Since the majority of those who use consultative decision-making are open to reach consensus during the consultation, there is a possibility of making collective decisions when there are no demanding constrains. Eleven out of fifteen software teams use either consultative or consensus decision-making styles, thus it is possible to claim that collaborative way of decision-making has a strong presence in the case company.

Despite the availability of various architecture decision-making techniques, none of the teams use any standard technique to make architecture decisions. Although a few teams use software tools to create diagrams that can be used for decision making and communication, the whiteboard is the standard tool for architecture decision making in the case company. Despite being an external entity, the majority of interviewees view architecture steering groups as useful bodies that support them in decision making. One of the main reasons given for this view is that these groups support the teams by reducing the complexity of the decision problem. Most of the time, software teams or their representatives take the initiative to consult the steering group. That can also have an impact on the teams' view on steering groups, as consulting the steering group is voluntary rather than forced upon the team.

## 5   Identified Architecture Decision-making Challenges

The interviewees mentioned several challenges associated with architecture decision making. Multiple interviewees provided evidence of the presence of *groupthink*, which leads groups to make inadequate decisions because it prevents them from taking actions required for informed decision making, including considering all possible alternatives, evaluating risks, examining decision objectives and seeking information related to the decision problem [17]. Based on the origin, the challenges are classified into three different groups: organisational, process and human. Table 3 shows identified challenges and their impact on architecture decision-making.

**Table 3.** Identified challenges and their impact on decision making

| Category | Challenge | Impact on Architecture Decision Making |
|---|---|---|
| Organisational | Inter team dependencies | Increased complexity |
| | Change of personnel | Loss of architecture knowledge |
| | Imposed technical constraints | Limit potential solutions |
| | Globally distributed teams | Lack of involvement |
| | Lack of a common tool chain | Difficult to collaborate |
| Process | Inadequate preparation time | Low quality decisions |
| | Dynamic requirements | Short term decisions |
| | Requirement ambiguity | Unclear decision goals |
| | Improper documentation | Missing design rationale |
| Human | Clash of personalities | Lengthy decision sessions / deadlocks |
| | Passive participation | Limited view points |

Revisiting the design rationale appears to be a significant problem due to improper documentation and organisational changes. Most of the interviewees admitted that they experience several issues related to the design documents, particularly regarding their quality and maintenance. The majority opinion is that the documentation practices in the company are minimal or, in some cases, non-existent. Multiple interviewees stated that differing opinions and personality traits are among the major challenges faced during architecture decision making. Several reasons such as non-flexibility, personal ego and loyalty towards a preferred technology prevent the team from reaching an agreement. Some team members constantly try to force their way of doing things on others rather than objectively participating in the discussion. On the other hand, some of the members prefer to just attend decision meetings but never express their opinions.

## 5 Conclusion

The study revealed that the majority of software teams in the company use a consultative decision-making approach to make architecture decisions. We were able to identify the challenges related from three different aspects: organisational, process and human, and their impact on architecture decision making. While discussing the overall results, we also uncovered the existence *groupthink* that is known to influence group decision making activities. The next logical step is to identify the relationship between the type of architecture decisions and the decision-making style followed. Identifying decision-making patterns that should be applied in different contexts will help software architects and teams select the best possible course of action to make their decisions. We are currently planning to cross analyse our previous case study findings [20] with the findings of this study to assess the generalisability.

**Acknowledgements.** This research is funded by ITEA2 and Tekes, the Finnish Funding Agency for Innovation, via the MERgE project, which we gratefully acknowledge. We would also like to thank all the interviewees and the management of the case company.

# References

1. Taylor, R.N., Medvidovic, N., Dashofy, E.M.: Software Architecture: Foundations, Theory, and Practice. John Wiley & Sons (2010).
2. Bass, L., Clements, P., Kazman, R.: Software Architecture in Practice. Addison-Wesley Professional (2012).
3. Clements, P.: A survey of architecture description languages. Proc. 8th Int. Work. Softw. Specif. Des. 16–25 (1996).
4. Jansen, A., Bosch, J.: Software Architecture as a Set of Architectural Design Decisions. In: 5th Working IEEE/IFIP Conference on Software Architecture (WICSA'05). pp. 109–120 (2005).
5. Kruchten, P.: Software architecture and agile software development: a clash of two cultures? 2010 ACM/IEEE 32nd Int. Conf. Softw. Eng. 2, 497–498 (2010).
6. Shore, J.: Continuous design. Software, IEEE. 21, 20–22 (2004).
7. Abrahamsson, P., Ali Babar, M., Kruchten, P.: Agility and Architecture: Can They Coexist? IEEE Softw. 27, 16–22 (2010).
8. Tofan, D., Galster, M., Avgeriou, P.: Difficulty of Architectural Decisions – A Survey with Professional Architects. In: Proceedings of 7th European Conference on Software Ar. pp. 192–199 (2013).
9. Rekhav, V.S., Muccini, H.: A Study on Group Decision-Making in Software Architecture. 2014 IEEE/IFIP Conf. Softw. Archit. 185–194 (2014).
10. Tofan, D., Galster, M., Lytra, I., Avgeriou, P., Zdun, U., Fouche, M.-A., de Boer, R., Solms, F.: Empirical evaluation of a process to increase consensus in group architectural decision making. Inf. Softw. Technol. 72, 31–47 (2016).
11. Rekha V, S., Muccini, H.: Suitability of Software Architecture Decision Making Methods for Group Decisions. 17–32 (2014).
12. Beebe, S.A., Masterson, J.T.: Communication in small groups: Principles and practice. Pearson Education Inc (2009).
13. Hersey, P., Blanchard, K.H., Johnson, D.E.: Management of Organizational Behavior. Pearson (2012).
14. Tannenbaum, R., Schmidt, W.H.: How to Choose a Leadership Pattern. Harv. Bus. Rev. 36, 95–101 (1958).
15. Stewart, L.P., Gudykunst, W.B., Ting-Toomey, S., Nishida, T.: The effects of decision-making style on openness and satisfaction within Japanese organizations. Commun. Monogr. 53, 236–251 (1986).
16. Schachter, S., Singer, J.E.: Assets and liabilities in group problem solving: the need for an integrative function. Psychol. Rev. 74, 239–249 (1967).
17. Janis, I.L.: Groupthink: Psychological Studies of Policy Decisions and Fiascoes. Cengage Learning (1982).
18. Eisenhardt, K.M.: Building Theories from Case Study Research. Acad. Manag. Rev. 14, 532–550 (1989).
19. Fantino, E., Stolarz-Fantino, S.: Decision-making: Context matters. Behav. Processes. 69, 165–171 (2005).
20. Dasanayake, S., Markkula, J., Aaramaa, S., Oivo, M.: Software Architecture Decision-Making Practices and Challenges: An Industrial Case Study. In: Proceedings of 24th Australasian Software Engineering Conference (2015).